# A Theory of Pragmatic Information and Its Application to the Quasi-species Model of Biological Evolution

Edward D Weinberger

*Max Planck Institute for Biophysical Chemistry, Am Fassberg, D-3400, Göttingen, Federal Republic of Germany*

Present address: Blumenthal Associates, Inc., 370 Central Park West, Suite 110, New York, NY, USA, 10025; Email: edw@BlumenthalAssociates.com.

## Abstract

**"Standard" information theory says nothing about the semantic content of information. Nevertheless, applications such as evolutionary theory demand consideration of precisely this aspect of information, a need that has motivated a largely unsuccessful search for a suitable measure of an "amount of meaning". This paper represents an attempt to move beyond this impasse, based on the observation that the meaning of a message can only be understood relative to its receiver. Positing that the semantic value of information is its usefulness in making an informed decision, we define *pragmatic information* as the information gain in the probability distributions of the receiver's actions, both before and after receipt of a message in some pre-defined ensemble. We then prove rigorously that our definition is the only one that satisfies obvious *desiderata*, such as the additivity of information from logically independent messages. This definition, when applied to the information "learned" by the time evolution of a process, defies the intuitions of the few previous researchers thinking along these lines by being monotonic in the uncertainty that remains after receipt of the message, but non-monotonic in the Shannon entropy of the input ensemble. It follows that the pragmatic information of the genetic "messages" in an evolving population is a global Lyapunov function for Eigen's quasi-species model of biological evolution. A concluding section argues that a theory such as ours must explicitly acknowledge purposeful action, or "agency", in such diverse fields as evolutionary theory and finance.**

## Introduction

By now, it is a cliché that the usual (Shannon) measure of information refers merely to the reduction in uncertainty resulting from the receipt of a message, and not to the *meaning* that the uncertainty reduction has to the receiver. Making this dichotomy between the *amount* of information contained in a message and the meaning of that information was, in some ways, a conceptual advance, because Shannon was interested in communication *per se*, as opposed to the content being communicated. Nevertheless, many authors have suggested that a theory that totally ignored semantics was, in some sense, incomplete.

For example, Warren Weaver's introduction to Shannon's seminal paper (Shannon and Weaver, 1962) observes that the effectiveness of a communications process could be measured by answering any of the following three questions

A. How accurately can the symbols that encode the message be transmitted ("the technical problem")?
B. How precisely do the transmitted symbols convey the desired meaning("the semantics problem")?
C. How effective is the received message in changing conduct ("the effectiveness problem")?

Weaver then observes that Shannon's paper --- and thus the entire edifice of what is now known as "information theory" --- concerns itself only with the answer to question A. The present paper, in contrast, attempts to address question C.

Workers in evolutionary theory feel the need for a theory of "information semantics" particularly acutely. Eigen (1971), in the same paper that introduces the quasi-species model, voiced the frustration of many authors when he wrote

> *Information theory as we understand it today is more a* communication theory. *It deals with problems of processing information rather than of "generating" information. It requires information to be present* "ab initio" *in a well defined form; …It always requires "somebody" -- usually man -- to determine what to call "information" and what to call "nonsense."*

and, later in the same paper,

> *This complementarity between information and entropy shows clearly the limited application of classical information theory to problems of evolution…. It is of little help as long as information has not yet reached its "full meaning", or as long as there are still many choices for generating new information. Here we need a new variable, a "value" parameter, which characterizes the level of evolution.*

Previous researchers have attempted to fill this need by assuming that the requisite measure of "amount of meaningful information" was equivalent to the "complexity" of a dataset. In other words, they have confused Weaver's Level B with Level C. The result is that, in the fifty or so years since Weaver's comments above, researchers still cannot agree upon a definition, let alone a theory of complexity. (For a critical review, see Atmanspacher and Weidenmann, 1999. Other critical comments in Standish, 2001 make clear that little had changed in the intervening two years. ).

Nevertheless, some of the results of that research can serve as a useful introduction to some of the ideas to be presented here. Of particular note was Grassberger's observation (Grassberger, 1986; Grassberger, 1991) that a low measure of meaningful information can result either from regularity or from randomness --- just as the regular object admits a simple description because of its regularity, so the random object admits a simple description as a member of a simply described ensemble. Expanding on that theme, Crutchfield and Young (1989) proposed an abstract notion of complexity "based on the information contained in the minimum number of equivalence classes obtained by reducing a data set modulo a symmetry". This definition offers the natural and generic dichotomy of macroscopic descriptions, which are given modulo the symmetry, and microscopic descriptions, which aren't. Furthermore, Crutchfield (1991) subsequently showed that, under rather general conditions, the complexity of the data set modulo the symmetry has a single maximum relative to the complexity of the data set without the symmetry. The high microscopic complexity regime then corresponds to "random" data in which the best macroscopic description is statistical. Unfortunately, for Crutchfield, and, more recently, for Standish (Standish, 2001), the relevant symmetry must be included explicitly in the analysis from the beginning, which seems to beg the fundamental question of how such symmetries should be identified in the first place.

The conduct-changing, or "pragmatic" import of a message is distinct from complexity in Crutchfield's sense because the receiver can choose to ignore any structure in the messages that is deemed irrelevant. Yet, judging from such reviews as Atmanspacher and Weidenmann (1999), this "pragmatic information" appears to have received little explicit attention. About the only reference that we have been able to find on the subject after repeated web searches is Atmanspacher and Scheingraber (1991), the reference that inspired the present work. These authors stop short of an explicit, quantitative definition of pragmatic information, but they do suggest that pragmatic information, like Crutchfield's complexity, must tend to zero in the extreme cases of "complete confirmation" and "complete novelty". In the first case, the messages merely confirm previous knowledge, obviating any need to change behavior. The uncertainty of what is to be sent, and thus the Shannon entropy of the underlying message ensemble tends to zero. In the second case, in which the messages contain material completely unrelated to that which has come before, the messages cannot be understood, let alone acted upon, because there is no context.

A fundamental assumption of the present work is that the practical meaning of information stems from its usefulness in making an informed decision. One important implication of this claim is a natural, quantitative measure of pragmatic information, which is the impact of a message on the receiver's subsequent actions. In contrast to the multitude of plausible definitions of complexity, we show that the pragmatic information measure is unique, up to an arbitrary choice of units, given certain natural assumptions. However, as Shannon also said about entropy, "the real justification of these definitions, … will reside in their implications."

One of the implications of this definition of pragmatic information is at least some of its hypothesized behavior in the limits described above. The limit of complete confirmation is established rigorously by Lemma 2, but, in the limit of complete novelty, we can show only that the pragmatic information has a fixed bound. In the special case of process, though, we show that this bound is indeed zero.

While the primary thrust of this paper is to introduce the mathematical theory, we demonstrate that the theory itself has ``pragmatic information" by demonstrating --- with the same model that Eigen used to elaborate the need for such a theory --- that pragmatic information does indeed answer Eigen's call for a "value parameter, [for] the level of evolution." Thus, the primary contributions of this paper are threefold:

1. A quantitative definition of pragmatic information in a general setting that is unique in satisfying certain plausible postulates, as well as being demonstrably correct in the previously mentioned limits.
2. A demonstration that pragmatic information is a global Lyapunov function for the quasi-species dynamics, regardless of initial conditions. This result can be viewed as an analogue of Boltzmann's celebrated *H*-Theorem of statistical mechanics (see, *e.g.* Thompson, 1972), at least for the "information generating" dynamics of the quasi-species. In other words, it is precisely pragmatic information that *is* being generated here.
3. The identification of the infinite time limit of the pragmatic information of the quasi-species as an important and potentially measurable parameter in evolutionary theory. As such, it is a more refined version of the "genomic complexity" introduced by Adami, *et. al.* (2000) to measure the growing number of conserved sites in the successful genomes of *in silico* selection experiments. Presumably, as the corresponding phenotypes "learn" about the environment, they act as the "receiver" for "messages" regarding which genomes are viable there. Under conditions to be specified below, the pragmatic information is the resulting entropy of the genome distribution. When the conditions are satisfied, the theorems stated above imply that the ability of the environment to ``compute'' evolutionary fitness places quantitative limits on the amount that natural selection can change the entropy of the genome.

The final section of the paper offers a context in which these results should be viewed.

## *A Formal Definition of Pragmatic Information*

The simplest case is the one in which one of a finite number of alternatives is to be selected, and selected once and only once. In this setting, pragmatic information is the mutual information between the ensemble of messages that could have been sent and the outcomes resulting from these messages. A number of intuitively appealing properties of pragmatic information follow immediately from this definition. In particular, many of the properties of Shannon information are shared by pragmatic information, as well as the uniqueness result described above.

Suppose a decision maker, $\mathcal{D}$, in some currents state, *s*, must choose among a set $\{\alpha_1, \alpha_2, \ldots \alpha_M\}$ of alternatives, assumed to be finite only to avoid unnecessary technicalities. Suppose further that these choices lead to outcomes $o \in \mathcal{O} = \{o_1, o_2, \ldots, o_N\}$, some of which the decision maker finds preferable to others. Finally, suppose that $\mathcal{D}$'s decision-making ability is constrained by one or more of the following:

- the mapping from the $\alpha$'s to the $o$'s is not deterministic,
- $\mathcal{D}$ has imperfect information about which of the $o$'s is, in fact, the best,
- $\mathcal{D}$ is unable to process the available information optimally,
- $\mathcal{D}$ is unable to guarantee that the decision made is the one actually implemented, possibly because of "environmental noise", *e,*
- Etc.

We reflect this indeterminacy by assuming that $\mathcal{D}$'s decision would result in a selection from the alternatives with respective probabilities $\boldsymbol{q} = (q_1, q_2, \ldots, q_N)$.

$\mathcal{D}$ is then presented with a "message" $m$, which may be the output of some process. Even with these new data, the best that $\mathcal{D}$ can do is to update the respective selection probabilities to $\boldsymbol{p}_m = (p_{1|m}, p_{2|m}, \ldots, p_{N|m})$ in such a way that the probability of a superior outcome is increased. This framework, illustrated in Figure 1, suggests the following

**Definition.** The ***pragmatic information***, $I_\mathcal{M}(\boldsymbol{p}; \boldsymbol{q})$, of the message ensemble $\mathcal{M}$ is the information gain in going from $\boldsymbol{q}$ to $\boldsymbol{p}_m$, averaged over all messages $m \in \mathcal{M}$, i.e.

$$I_M(p; q) = \sum_{i,m} p_{i|m} \varphi_m \log_2\left(\frac{p_{i|m}}{q_i}\right)$$

$$= \sum_{i,m} p_{i,m} \log_2\left(\frac{p_{i,m}}{\varphi_m q_i}\right) \quad (1)$$

where $\varphi_m$ is the marginal probability that message $m$ was sent and $p_{i,m}$ is the joint probability that message $m$ was sent and outcome $\sigma_i$ was realized.

It will be assumed that, included in $\mathcal{M}$ is the "empty message", $\nu$, which, by definition, results in no change to the *a posteriori* probabilities. The presence or absence of the empty message in $\mathcal{M}$ makes no difference to the computation, as the terms involving $\nu$ evaluate to zero. However, the ability to write $q_i = p_{i|\nu}$ will eliminate potential confusion in the discussion of joint and conditional pragmatic information below.

(Figure 1 about here…)

Note also that Figure 1 is significantly different from the standard picture drawn in information theory textbooks, beginning with Shannon and Weaver (1962). In contrast to "standard information theory," which considers only the receiver's ability to distinguish the received message from others in the ensemble, the decision maker is central to the present work.

Our thinking here is informed by our experience in programming the guidance, navigation, and control system of the Space Shuttle project. After the guidance system identified a desired course for the Shuttle, and after the navigation system decided what the current course of the Shuttle actually was, the flight control system tried to address any discrepancies by sending signals to the rudder, flaps, etc. To use the terminology of Figure 1, the guidance and navigation systems are sources of messages to be received and acted upon by the "decision making" capabilities of the flight control system. However, the term "Decision Maker" as used in the Figure could also apply to an animal, a person, or even a number of people jointly making decisions. What is important here is the notion of *agency*, that the decision maker "effects" its decision by purposeful action, a point that we discuss in detail in the concluding section of the paper.

It is this attempt at purposeful action that justifies the asymmetry between $p_m$ and $q$ in the definition above. Presumably, the new distribution is a better predictor than the old one as to which outcomes serve the stated purpose. Also implicit in the definition --- through the requirement that $q_i > 0$ for all $i$ in order for the pragmatic information to be defined at all --- is the assumption that the messages cannot open up new possibilities, but merely alter the likelihoods of existing ones.

## *Some Basic Lemmas*

Some elementary, but important properties of pragmatic information are contained in the following lemmas.

**Lemma 1:** Pragmatic information can be written as the mutual information of $\mathcal{M}$ and $O$, which can assume any one of three equivalent characterizations in terms of the (Shannon) entropy (Shannon and Weaver, 1962).

Let
$$\mathcal{H}(\mathcal{M}) = -\sum_m \varphi_m \log_2 \varphi_m,$$
the entropy of the message ensemble $\mathcal{M}$, let
$$\mathcal{H}(O) = -\sum_i q_i \log_2 q_i,$$
the entropy of the outcome ensemble $O$, let
$$\mathcal{H}(O|\mathcal{M}) = -\sum_{i,m} p_{i,m} \log_2 p_{i|m},$$
the conditional entropy of the outcomes, given the various messages, and let
$$\mathcal{H}(\mathcal{M}|O) = -\sum_{i,m} p_{i,m} \log_2 p_{m|i},$$
the conditional entropy of the messages, given a particular outcome. Then
$$\begin{aligned}\mathcal{I}_\mathcal{M}(p;q) &= \mathcal{H}(\mathcal{M}) + \mathcal{H}(O) - \mathcal{H}(O,\mathcal{M}) \\ &= \mathcal{H}(O) - \mathcal{H}(O|\mathcal{M}) \\ &= \mathcal{H}(\mathcal{M}) - \mathcal{H}(\mathcal{M}|O)\end{aligned}$$

Because $\mathcal{H}(\mathcal{M}|O) \geq 0$, it follows immediately from the third identity that

**Lemma 2:** The pragmatic information of a message ensemble is always bounded above by its Shannon entropy.

Ref. 1 also proves, albeit in a different context, that

**Lemma 3:** $\mathcal{I}_\mathcal{M}(p;q)$ is invariant with respect to coordinate transformations and automorphisms of the outcomes. In other words, $\mathcal{I}_\mathcal{M}(p;q)$ is dimensionless.

**Lemma 4:** $\mathcal{I}_\mathcal{M}(p;q) \geq 0$, with equality only when $q$ and $p_m$ coincide for all $m$.

*Proof:* The only extremum of $\mathcal{I}_\mathcal{M}(p;q)$ subject to $\sum_i p_{i|m} = 1$ and $\sum_i q_i = 1$ found via Lagrange multipliers is the value $\mathcal{I}_\mathcal{M}(p;q) = 0$ when $q$ and $p_m$ coincide for all $m$. It therefore suffices to check a single case where $q$ and $p_m$ do not coincide to determine whether zero is a minimum or a maximum. Choosing $p_{1|m} = 1$ and $p_{i|m} = 0$ for $i > 1$, we see that $I_M(p; q) > 0$ when $q$ and $p_m$ do not coincide. ∎

Lemma 4 implies the intuitively reasonable result that no pragmatic information accrues from processes in which the probabilities of the outcomes are preserved. In particular, $I_M(p; q) = 0$ unless the inputs to the decision process affect the computations underlying $\mathcal{D}$'s response. Considerations such as this suggest the importance of the computational abilities of the receiver, but that is a subject for future research.

**Lemma 5:** For all fixed $q$, $I_M(p; q)$ is a local maximum with $p_m$ lying on the vertices of the simplex $p_{i|m} \geq 0$, $\sum_i p_{i|m} = 1$. Furthermore, there are no local maxima in the interior of the simplex (In other words, the maximal amount of pragmatic information accrues to messages that engender complete certainty, independent of the Shannon entropy of the message ensemble.).

*Proof:* For all $p_m = (p_{1|m}, p_{2|m}, \ldots, p_{N|m})$ on the unit simplex,

$$I_M(q) = \sum_{i<N} p_{i|m} \log_2 p_{i|m}/q_i + \left(1 - \sum_{i<N} p_{i|m}\right) \log_2 \left(1 - \sum_{i<N} p_{i|m}\right)/q_N.$$

Taking the partial derivative with respect to $p_{j|m}$, with $j < N$, we have

$$\partial I_M(q) / \partial p_{j|m} = \log_2 [p_{j|m}/(1 - \sum_{i<N} p_{i|m})] / \ln 2 + \log_2 [q_N/q_j] / \ln 2,$$

which is monotone increasing in $p_{j|m}$. Therefore, by the chain rule, the directional derivative along any ray that extends through one of the verticies of the simplex into its interior must also be monotone increasing. Because any interior maxima would have to be connected to one of the verticies by one such ray, along which $I_M$ must increase until the intersection with the putative maximum, and then decrease, such maxima cannot exist. Furthermore, for small $p_{j|m}$, i.e. in the neighborhood of $e_j$, the derivatives are large and negative, implying that small departures from $e_j$ decrease $I_M$, so $e_j$ is a local maximum. The $p_{j|m}$'s can clearly be redefined so that this last argument applies to any of the verticies of the simplex. ∎

**Lemma 6:** $$I_M(p; q) \leq \max_i \left[-\log_2 q_i\right]$$

*Proof:* Follows immediately from the definition and Lemma 5. ∎

### *Joint and Conditional Pragmatic Information*

We define $I_{M,N}(p; q)$, the joint pragmatic information of messages $m \in M$ and $n \in N$, as

$$I_{M,N}(p; q) = \sum_{i,m,n} p_{i,m,n} \log_2 \left(\frac{p_{i|m,n}}{q_i}\right), \qquad (2)$$

where $p_{i,m,n}$ is the joint probability that both messages $m$ and $n$ were sent and outcome $\sigma_i$ was realized and $p_i|_{m,n}$ is the corresponding conditional probability. Similarly, the conditional pragmatic information of messages $m \in M$, given $n \in N$ is

**Lemma 7:** $$I_{M,N}(p; q) = I_{M|N}(p; q) + I_N(p; q)$$

*Proof:*

$$\begin{aligned}
I_{M,N}(p; q) &= \sum_{i,m,n} p_{i,m,n} \log_2 \left( \frac{p_{i|m,n}}{p_{i|n}} \frac{p_{i|n}}{q_i} \right) \\
&= \sum_{i,m,n} p_{i,m,n} \log_2 \left( \frac{p_{i|m,n}}{p_{i|n}} \right) + \sum_{i,m,n} p_{i,m,n} \log_2 \left( \frac{p_{i|n}}{q_i} \right) \\
&= \sum_{i,m,n} p_{i,m,n} \log_2 \left( \frac{p_{i|m,n}}{p_{i|n}} \right) + \sum_{i,n} p_{i,n} \log_2 \left( \frac{p_{i|n}}{q_i} \right) \\
&= I_{M|N}(q) + I_N(q)
\end{aligned}$$

∎

The receiver of the messages can obviously make more inferences given both $m \in M$ and $n \in N$ than from either $m$ or $n$ by themselves. Hence, our definition of pragmatic information is consistent with a postulate of "semantic information" of Bar Hillel and Carnap (1953), who require that the semantic information of a message increase with the number of deductions that follow from the message. This requirement, that

$$I_{M,N}(p; q) \geq I_M(p; q)$$

is an immediate consequence of Lemma 7 and the fact (Lemma 4) that $I_{M|N}(p; q) \geq 0$.

Furthermore, if $m$ can always be inferred from $n$, then $I_{M|N}(p; q) = 0$. This condition is one of "pragmatic sufficiency", named to suggest a *sufficient statistic*, knowledge of which is "sufficient" to answer a question of interest (See, for example, Silvey, 1975).

We also have $I_{M|N}(p; q) = I_M(p; q)$ ("pragmatic independence") of message ensembles $M$ and $N$ if the equivalent conditions

$$p_{i|m,n} / p_{i|n} = p_{i|m} / p_{i|v} \quad \Leftrightarrow \quad p_{i|m,n} / p_{i|m} = p_{i|n} / p_{i|v} \quad (3)$$

hold for all $m \in M$ and $n \in N$ (Recall that the symbol $p_{i|v}$, i.e. the probability of outcome $\sigma_i$ when the null message has been sent, is another way of writing $q_i$, the *a priori* probability of outcome $\sigma_i$.) (3) implies that no $n \in N$ can influence the processing of any $m \in M$ and *vice versa*. These conditions hold if $p_{i,m,n} = p_{i|v} \varphi_m \varphi_n$ for all values of these subscripts, but

**Lemma 8:** The pragmatic independence of message ensembles $M$ and $N$ need not imply that all pairs $m \in M$ and $n \in N$ are independent events in the standard probabilistic sense. It is also possible for $M$ and $N$ to be pragmatically dependent and $m \in M$ and $n \in N$ to be probabilistically independent events for all pairs $m \in M$ and $n \in N$.

*Proof:* Writing (3) using joint, rather than conditional probabilities, we have

$$p_{imn} = \frac{\varphi_{mn}}{\varphi_m \varphi_n} \frac{p_{im} p_{in}}{p_{i|\nu}} . \qquad (4)$$

If $i$, $m$, and $n$ can all assume the values 0 and 1, and

$$p_{imn} = \frac{1}{8} + (-1)^{\delta_{im}} a + (-1)^{\delta_{in}} b$$

for $|a| \leq 1/8$ and $|b| \leq 1/8 - |a|$, then $\varphi_{mn} = 1/4$ for all $m$ and $n$, and $\varphi_m = \varphi_n = 1/2$ for all $m$ and $n$. Thus, $\varphi_m \varphi_n = \varphi_{mn}$ for all $m$ and $n$, but the condition for pragmatic independence does not hold. Conversely, any family of probabilities chosen such that $\varphi_m \varphi_n \neq \varphi_{mn}$, for some $m$ and $n$, but $p_{i,m} = p_{i|\nu} \varphi_m$, $p_{i,n} = p_{i|\nu} \varphi_n$, and $p_{i,m,n} = p_{i|\nu} \varphi_{mn}$ for all $m$ and $n$, will satisfy (4). ∎

## *The Uniqueness Theorem*

A fundamental property of Shannon information is the behavior of the information measure when the choice among the outcomes is broken down into two or more successive choices (Shannon and Weaver, 1962). To consider the same situation for pragmatic information, let $N$ be the number of elements of $O$, and, *a priori*, assign probability $1/N$ to each $o \in O$. Assume further that $O$ is partitioned into the $K < N$ outcome classes $O_1, O_2, \ldots, O_k \ldots, O_K$, with $N_1, N_2, \ldots, N_k, \ldots N_K$ outcomes in the respective classes.

Definition (1) assigns pragmatic information $I_j = \log_2(N/N_j)$ to any message in an ensemble that unambiguously eliminates all but the members of the $j^{th}$ class from consideration. Less definitive messages might not provide sufficient grounds to eliminate the other classes completely; instead, such messages might change the *a posteriori* class probabilities from $N_j/N$ to $p_j > 0$ for more than one class. In that case, it might seem reasonable that the pragmatic information of such a message is the weighted sum of the $I_j$'s. However, the $I_j$'s were computed by assuming more information than is available; namely, the certainty that outcomes outside of the $j^{th}$ class had already been eliminated from consideration. Definition (1) therefore subtracts from the weighted sum of the $I_j$'s the Shannon information required to distinguish with certainty *which* class is not eliminated.

The above considerations are sufficient to determine the functional form of pragmatic information to within a constant factor, which is effectively a choice of units. This fact is the content of

**Theorem 1.** $I_M(p; q)$ as defined above is the unique continuous function of the $2N$ arguments, $p = (p_1, p_2, \ldots, p_N)$ and $q = (q_1, q_2, \ldots, q_N)$ such that

i. If $q_i = 1/N$ for all $N$ outcomes $o \in O$, and $p_i = 1/P$ for $o_i \in O^* \subset O$, where $O^*$ has $P < N$ elements, and $p_i = 0$ for all other $i$, then $I_M(p; q)$ is monotone in $P/N$.

ii. $I_M(p; q)$ is additive for pragmatically independent messages.

iii. If the outcome ensemble is partitioned into $K$ sub-ensembles, $O_1, O_2, \ldots, O_K \subset O$, with $N_1, N_2, \ldots, N_K$, outcomes respectively, and $I_1, I_2, \ldots, I_K$ are the respective pragmatic information values when, respectively, $O_1, O_2, \ldots,$ or $O_K$ contain $o_i$ with certainty, then, for general $p_{j|m}$, the overall pragmatic information is the weighted sum

$$I_M(p; q) = \sum_{m \in M, 1 \leq j \leq K} p_{j,m} I_j - \mathcal{H}(O | M),$$

where those symbols defined previously have the same meaning as before.

*Proof*: See Appendix I.

### *The Pragmatic Information of a Process*

Pragmatic information can also accrue from the time evolution of a process, if measurement of the process's state is used to update *a priori* predictions. This is really a special case of the above in which measurement(s) at various times are the only "messages" received. For example, we might start with the prediction that the probability, $p_k(t_2)$, of observing the process in state $k$ at a given time $t_2$ is simply the same as the probability, $p_k(t_1)$, of this observation at a previous time $t_1$. The pragmatic information of the observation at $t_2$ is then

$$\sum_k p_k(t_2) \log_2[p_k(t_2)/p_k(t_1)].$$

Alternatively, we might use the observed state, *i*, of the process at the previous time to predict the probability of state *k* at the subsequent time. Assuming that both observations are to be made in the future, the pragmatic information of learning state *k* at the subsequent time, given the observation of state *i* at the previous time is

$$\sum_{i,k} p_{i,k}(t_1, t_2) \log_2 \left[ \frac{p_{i,k}(t_1, t_2)}{p_i(t_1) p_i(t_1)} \right],$$

where we denote the joint probabilities of both observations by $p_{i,k}(t_1, t_2)$. In the limit of complete novelty (maximum Shannon entropy) discussed above, we have $p_i(t_1) = p_k(t_1) = p_k(t_2)$, for all states $i$ and $k$, and for all times $t_1$ and $t_2$. Since we also have $p_{i,k}(t_1, t_2) = p_i(t_1) p_k(t_2)$ in that case (Otherwise, the conditional entropy of the state at $t_2$, given the state at $t_1$ will not be a maximum.), it follows that

**Lemma 9**: The pragmatic information of a process is zero in the limit of complete novelty.

We note the important role that additivity plays in the above.

In the next section, the occupation probabilities will represent the probabilities of choosing an individual of each of the various types in a random sample of the population. It is reasonable to assume that each such sample is independent of previous samples, in which case the joint pragmatic information formula simplifies to

$$\sum_k p_k(t) \log_2 \left[ \frac{p_k(t)}{p_k(0)} \right]. \tag{5}$$

### *Application to the Quasi-species Model*

As an example of how pragmatic information can be applied, we return to the problem of evolution, which we now restate more explicitly as one of characterizing the information-gathering activity of the so-called quasi-species model (Eigen, 1971; Eigen and Schuster, 1979; Eigen, M$^c$Caskill, and Schuster, 1988). For present purposes, the model can be viewed as a simplified model of Darwinian selection with mutation. Here, the environment effectively "decides" on the fitness of the phenotype corresponding to each given genotype, where a phenotype's fitness is defined to be its reproduction rate. The corresponding genotypes are the messages that inform the decision by determining the composition of subsequent generations.

Given suitable assumptions (Eigen, 1971; Eigen and Schuster, 1979; Eigen, M$^c$Caskill, and Schuster, 1988), the quasi-species model describes, in the large population (continuous) limit, the relative frequencies $\boldsymbol{x} = (x_1, x_2, \ldots x_i, \ldots)$ of abstract "replicators" subject to differential selection and mutation according to the ordinary differential equation

$$d\boldsymbol{x}(t)/dt = \mathbf{W}\boldsymbol{x}(t) - \boldsymbol{x}(t)\left[\mathbf{1}^T \mathbf{W} \boldsymbol{x}(t)\right] = \left[\mathbf{W} - \Omega(t)\right]\boldsymbol{x}(t), \tag{6}$$

satisfying the constraint that $\mathbf{1}^T \boldsymbol{x}(t) = 1$ for all times, $t$. Here, $\mathbf{1}$ is a column vector of all 1's, $\mathbf{W}$ is a square matrix whose diagonal elements $W_{ii}$ represent the rates at which replicators of type $i$ are copied successfully, and whose off-diagonal elements, $W_{ij}$, represent the rate at which type $j$ is produced because of mutations while copying type $i$, $\left[\mathbf{1}^T \mathbf{W} \boldsymbol{x}(t)\right] = \Omega(t)$, and $^T$ is the transpose operation. It is usually assumed that $\mathbf{W} = \mathbf{RQ}$, where $\mathbf{R}$ is a diagonal matrix of replication rates, and $\mathbf{Q}$ is a symmetric matrix whose components $Q_{ij}$ specify the probabilities of errors that result in a copy of species $j$ instead of species $i$ (Eigen, M$^c$Caskill, and Schuster, 1988). These same authors cite references showing that $\mathbf{W} = \mathbf{RQ}$ has real eigenvalues. In any case, we have

$$x(t) = e^{\mathbf{W}t} x(0) \;/\; [\mathbf{1}^T e^{\mathbf{W}t} x(0)] = \exp[\mathbf{W}t - \int_0^t \Omega(s)\,ds]\, x(0), \qquad (7)$$

where $x_i(0)$ is the population of each type of replicator at some initial time. If $W_{ij} > 0$ for all $i$ and $j$, and $W_1 > W_i$ for all $i > 1$, then

$$\lim_{t \to \infty} x_i(t) \approx \begin{cases} c, & \text{if } i = 1, \\ c\,[W_{1i}/(W_{11} - W_{ii})] & \text{otherwise,} \end{cases} \qquad (8)$$

where the normalization constant $c = \left[1 + \sum_{i>1} W_{1i}/(W_{11} - W_{ii})\right]^{-1}$, independent of the initial conditions (Eigen, M$^c$Caskill, and Schuster, 1988). Typically, $W_{1i}/(W_1 - W_i) \ll 1$, for all $i > 1$, so it appears that "natural selection" has acted on the system to identify the "fittest" replicator type and thus "generate" information. Upon closer examination, though, the nature of this information proves elusive, as the same number of binary bits would be required to specify *any* particular sequence of the same length, regardless of its fitness. Furthermore, the same kind of process is taking place, regardless of whether the initial conditions are

- equal numbers of each different type of replicator,
- a population consisting only of replicators of type 1, or
- a population consisting only of replicators of some inferior type.

Pragmatic information provides a meaningful measurement of the rate at which this process proceeds in each of these cases. At each time $t$, the environment receives "messages" about the fitness of a particular replicator via the number of copies of that replicator's genome. Prior to receipt of the messages (and the subsequent "processing" that results in differential selection), the initial probability of selecting a replicator of type $i$ at random from the system is $q_i = x_i(0)$. The probability of selecting a replicator of type $i$ at subsequent time $t$ is $p_i = x_i(t)$. That the pragmatic information provides a direction to quasi-species evolution, much as the Second Law of Thermodynamics directs thermodynamic processes, is the content of

**Theorem 2.** Pragmatic information is a global Lyapunov function for the quasi-species model if $\mathbf{W} = \mathbf{RQ}$, where $\mathbf{R}$ is a diagonal matrix with diagonal entries $R_{ii} \neq 0$ and where $\mathbf{Q}$ is symmetric. In other words, the rate at which pragmatic information is generated by the quasi-species model if $\mathbf{W} = \mathbf{RQ}$ is always positive for finite time, regardless of initial conditions, and it tends to zero in the infinite time limit.

*Proof:* See Appendix II.

A plot of $x(t)$ and $\mathcal{I}(x(t); x(0))$ in a typical case in which $x^T(0) = (0.49, 0.005, 0.49, 0.005)$ in arbitrary units and

$$\mathbf{W} = \begin{bmatrix} 1.7755 & 0.0181 & 0.0181 & 0.0002 \\ 0.0625 & 6.1210 & 0.0006 & 0.0625 \\ 0.0268 & 0.0003 & 2.6243 & 0.0268 \\ 0.0005 & 0.0501 & 0.0501 & 4.9051 \end{bmatrix},$$

chosen to exhibit a variety of the possible behaviors of the $x_i$'s, is shown in Figure 2, below.

(Figure 2 about here…)

For non-zero mutation rates, Lemma 5 implies that

$$\lim_{t \to \infty} I(x(t); x(0)) \approx -c \log_2 x_1(0) + \sum_{i>1} c \log_2 [cW_{1i}/x_i(0)(W_{11} - W_{ii})] \quad (9)$$

$$\leq -\log_2 x_1(0). \quad (10)$$

Note that equality in (10) is achieved only when the mutation rate is zero. For non-zero mutation rates, the optimizing ability of natural selection is limited by the "pragmatic noise" introduced into the selection process by random mutation. The infinite time limit in (9), which monotonically decreases with an increasing average mutation rate, can thus be used as a natural measure of the relative strength of selective forces relative to mutation.

Note also that in the limit of zero mutation, the solution to the quasi-species equation is a Boltzmann distribution with a temperature parameter that decreases as time progresses, establishing a direct physical correspondence between the quasi-species equation and a system cooling in thermodynamic equilibrium. In fact, the well-known characterization of the Boltzmann distribution as that distribution with a given mean and maximum entropy (see, *e.g.* Haken, 1983) generalizes as follows:

**Lemma 10:** For a given prior distribution **q** and a given value of

$$<W> = \sum_i W_{ii}\, p_{i|m},$$

the distribution $p_{i|m}$ that maximizes the pragmatic information given message $m$ is

$$p_{i|m} = q_i \exp(\mu W_{ii}) / [\sum_k q_k \exp(\mu W_{kk})]$$

where $\mu$ is that unique value for which

$$<W> = [\sum_k q_k W_{kk} \exp(\mu W_{kk})] / [\sum_k q_k \exp(\mu W_{kk})]$$

*Proof:* The proof is a straightforward generalization of the Lagrange multipliers argument that justifies the result cited in Haken (1983). For given $m$, the pragmatic information,

$$I_m(q) = \sum_i p_{i|m} \log_2\left(\frac{p_{i|m}}{q_i}\right)$$

is maximized for the given constraints (including the constraint that the *a posteriori* probabilities $p_{i|m}$ sum to unity) is an extremum at extrema of the Lagrangian

$$L = \sum_i p_{i|m} \log_2\left(\frac{p_{i|m}}{q_i}\right) + \lambda\left(\sum_i p_{i|m} - 1\right) + \mu\left(\sum_i p_{i|m} W_{ii} - 1\right)$$

Upon setting the derivatives of $L$ with respect to $p_{k|m}$ for all $k$, $\lambda$, and $\mu$ to zero and solving simultaneously, the result follows. ∎

This result might be viewed as a justification for using the Boltzmann distribution in simulated annealing (see, for example, Kirkpatrick, *et. al.*, 1983, for a description of the algorithm.).

### *On the Notion of Agency*

Up until now, science has dismissed the notion of agency in, for example, biology, as "teleological". Nevertheless, Kauffman (2000) makes a compelling case for considering "autonomous agents" in a wide variety of disciplines, including economics and evolutionary theory. While Kauffman's agents can also reproduce and perform at least one thermodynamic cycle, their most fundamental property is precisely their ability to act purposefully to further their goals (Kauffman's endowment of his agents with the ability to reproduce and the need to perform thermodynamic cycles suggests the obvious goal of obtaining the resources required for reproduction and maintenance.).

Having shown that pragmatic information is an order parameter in measuring the "decisiveness" of the environment in the quasi-species model, we speculate that it might serve equally well in some of the other contexts that Kauffman considers. For example, the "efficient market hypothesis" that undergirds essentially all of modern finance is succinctly, but precisely stated as the claim that the pragmatic information of previous price histories is zero.

The best scientific theories seek to explain some range of phenomena in terms of a parsimonious set of principles that are "nearer to hand" than obscure, if fundamental truths. The price paid for this conceptual transparency is the approximation of complex microscopic behavior by single macroscopic quantities, such as friction, ferromagnetization, and the like. Unfortunately, these "order parameters" must be written into the theory "by hand;" in fact, identifying appropriate order parameters is often the crux of formulating the appropriate theory!

The present approach skirts this problem by asking a question that is often easier to address; namely, what aspects of the situation are determinants of behavior. This approach fits naturally into Rosen's "relational biology" (Rosen, 1991). Central to this vision is the concept of function, which, Rosen objectifies via the following question:

> *"Suppose ... we are given a system, or better, a state that is perceptibly heterogenious; one part looks different, or behaves differently from another part. If we leave the system alone, some autonomous behavior will ensue. On the other hand, we can ask ca question like: if we were to remove, or change, one of these distinguishable parts, what would be the effect on that behavior?"* (Italics in the original).

Clearly, the present paper is nothing but an answer to that question for the quasi-species.

The above quotation was taken from Rosen's monograph, *Life Itself* (Rosen, 1991), which makes a persuasive case that the reductionistic methods of contemporary science are inadequate to the task of understanding living organisms. Not coincidentally, this same monograph contains one of the few references in the modern scientific literature to Aristotle's Theory of Causation. Where modern science might seek *the* cause for a particular phenomenon, Aristotle would identify no less than *four* different kinds of causation, "formal cause", "material cause", "efficient cause", and "final cause". The first three of these coincide with contemporary scientific ideas of causation --- material causation corresponding to the initial conditions leading to a particular phenomenon, formal causation corresponding to the set of dynamical rules to be applied to the initial conditions, and efficient causation with the particular application of the rules to the particular initial conditions to produce the phenomenon in question. In contrast, modern science barely acknowledges final causation, the category Aristotle considered most important, because final causation implies purposeful action.

Historically, this reluctance to consider final causation was a step forward, because it directed attention towards observation and measurement, and away from sterile theological debate. Nevertheless, Rosen argues, biology cannot be properly understood without at least some consideration of final causation. In place of his book-length defense of this thesis, we close with the quotation from Rosen's mentor, Nicholas Rashevsky, that Rosen uses to preface his argument:

> *It is important to know how pressure waves are reflected in blood vessels. It is important to know that diffusion drag forces may produce cell division. It is important to have a mathematical theory of complicated neural networks. But nothing so far in these theories indicates that the proper functioning of arteries and veins is essential for the normal course of the intracellular processes; nor does anything in those theories indicate that a complex phenomenon in the central nervous system ... [is] tied up with metabolic processes of other cells in the organism.... And yet this integrated activity of the organism is probably the most essential manifestation of life.*

> *So far as the theories mentioned above are concerned, we may just as well treat, in fact,* do *treat, the effects of diffusion drag forces as a peculiar diffusion in a rather specialized physical system, and we do treat the problems of circulation as special hydrodynamic problems. The fundamental manifestation of life mentioned above drop out from all our theories in mathematical biology.*

> *...We must look for a principle which connects the different physical phenomena involved and expresses the biological unity of the organism and the organic world as a whole.*

The goal of the present paper, by making explicit the role of agency in information processing, is to help biology to move beyond the "mere syntax" of the "rather specialized physical system" towards a more "organic" view of life.

### *Acknowledgements*

This work was funded partially by a Max Planck Stipendium in] the laboratory of Prof. Manfred Eigen that was awarded to the author from 1989 to 1991. The award included travel to a NATO Conference on Information Dynamics in Irsee, Germany in 1991. The author wishes to acknowledge one of the organizers of that conference, Harald Atmanspacher, for useful discussions, Prof. Eigen for identifying the issues addressed by this paper and for providing a stimulating atmosphere for research, and the two anonymous referees for perceptive reviews of the manuscript.

Parts of this work were presented at a Santa Fe Institute workshop on adaptive landscapes held in July, 1995. The author wishes to acknowledge the ensuing discussion with Jim Crutchfield for helping to clarify the distinction between his work and the present work.

# *APPENDIX I: PROOF OF THEOREM 1*

As was noted above, messages are pragmatically independent if they are the concatenation of probabilistically independent component messages that refer to probabilistically independent sub-events. We consider the case in which there are $n$ such message component, sub-event pairs. In the spirit of condition *i*, we assume that the *a priori* probability of the outcome of each sub-event is $1/N$, that the *a posteriori* probability is $1/P$, and that the pragmatic information of each outcome is $A(N/P)$. The respective *a priori* and *a posteriori* probabilities are $1/N^n$ and $1/P^n$. Thus, by additivity, $A(N^n/P^n) = n\, A(N/P)$, for all $n$, $P$, and $N$. Because there exists choices of the integers $N$ and $P$ such that the ratio $N/P$ approximates any real number greater than or equal to unity, the continuity of $A$ guarantees that $A(t^n) = n\, A(t)$ for all real $t \geq 1$ and all $N$. This last result is sufficient to guarantee (Shannon and Weaver, 1962) that $A(t) = c \log t$, for some constant c and all real $t > 1$. The constant can be interpreted as a choice of units, or the choice of the base of the logarithm, which is effectively the same thing. As in the Shannon theory, it is natural to use base 2 logarithms.

Suppose next that the original $N$ *a priori* outcomes have been partitioned into $K$ classes, each of which has $N_j$ elements, for $1 \leq j \leq K$. If each outcome has the same *a priori* chance of being selected, the *a priori* probability that the selected outcome is in class $j$ is $N_j/N$. Furthermore, if the *a posteriori* probability of selecting an outcome from class $j$, given that message $m$ is received is $P_{j|m}$, then, by condition *iii* and the previous paragraph,

$$I_M(q) = \sum_{m \in M} \sum_{1 \leq j \leq K} \varphi_m\, p_{j|m} \log_2 (N/N_j) - \mathcal{H}(O \mid M),$$
$$= \sum_{m \in M} \sum_{1 \leq j \leq K} \varphi_m\, p_{j|m} \log_2 (p_{j|m} N/N_j)$$

Because values of $N_j$ and $N$ can be found that approximate any given probability $q_j$, the theorem follows. ∎

# APPENDIX II: PROOF OF THEOREM 2

Having demonstrated above that pragmatic information is zero if and only if the *a posteriori* and *a priori* distributions coincide, that it attains its unique maximum at one of the verticies of the unit simplex, and that $\lim_{t \to \infty} d\mathcal{I}(x(t); x(0))/dt = 0$, we need only show $d\mathcal{I}(x(t); x(0))/dt > 0$ for all finite times. We have

$$\frac{dI}{dt} = \sum_i \frac{dx_i(t)}{dt} \log_2 \left( \frac{x_i(t)}{x_i(0)} \right)$$

$$= \frac{1}{\ln 2} \int_0^t \sum_i \frac{dx_i(t)}{dt} \left[ \frac{1}{x_i(r)} \right] \frac{dx_i(r)}{dr} dr.$$

Using the matrix-vector form of the given differential equation to eliminate the derivatives in the sum above and the fact that $\Omega(t) - \Omega(r)$ is a multiple of the identity matrix, we also have

$$\frac{dI}{dt} = \frac{1}{\ln 2} x(t)^H [\mathbf{W} - \Omega(t)]^H \int_0^t \text{diag}\left[\frac{1}{x(r)}\right] [\mathbf{W} - \Omega(r)] x(r) dr$$

$$= \frac{1}{\ln 2} x(t)^H [\mathbf{W} - \Omega(t)]^H \int_0^t \text{diag}\left[\frac{1}{x(r)}\right] [\mathbf{W} - \Omega(t)] x(r) dr +$$

$$\frac{1}{\ln 2} x(t)^H [\mathbf{W} - \Omega(t)]^H \int_0^t \text{diag}\left[\frac{1}{x(r)}\right] [\Omega(t) - \Omega(r)] x(r) dr$$

$$= \frac{1}{\ln 2} x(t)^H [\mathbf{W} - \Omega(t)]^H \int_0^t \text{diag}\left[\frac{1}{x(r)}\right] [\mathbf{W} - \Omega(t)] x(r)$$

The second sum in the second line vanishes, because it is proportional to $\mathbf{1}^T dx(t)/dt = 0$ for all times, $t$. To verify that the remaining sum is positive, set $z = (t-r)/2$ and $s = (t+r)/2$, so that

$$x(t) = \exp\left[\mathbf{W}z - \int_s^t \Omega(s)\, ds\right] x(s),$$

and

$$x(r) = \exp\left[-\mathbf{W}z + \int_r^s \Omega(s)\, ds\right] x(s).$$

Thus, because $\exp[\pm \mathbf{W}z]$ commutes with $\mathbf{W}$ for all $z$, the theorem will follow if, for all vectors $x(s)$,

$$x^H(s)[\mathbf{W} - \Omega(t)]^H \left\{ \exp[\mathbf{W}z]^H \text{diag}\left[\frac{1}{x(r)}\right] \exp[-\mathbf{W}z] \right\} [\mathbf{W} - \Omega(t)] x(s) > 0,$$

or,

$$y^H \exp[\mathbf{W}z]^H \mathbf{D} \exp[-\mathbf{W}z]y > 0,$$

where $\mathbf{D}$ is a diagonal matrix with all diagonal elements positive and $y$ is any real vector of appropriate dimension. We exploit the special form of $\mathbf{W}$ to write

$$\mathbf{W} = \mathbf{RQ} = \mathbf{R}^{1/2}(\mathbf{R}^{1/2}\mathbf{T}^H)\Lambda(\mathbf{TR}^{1/2})\mathbf{R}^{-1/2},$$

where $\Lambda$ is the diagonal matrix of eigenvalues of $\mathbf{Q}$, and $\mathbf{T}$ is the unitary matrix of eigenvectors of $\mathbf{Q}$. Since $\mathbf{Q}$ is symmetric, $\Lambda$ must be real. Thus,

$$\exp(\mathbf{W}z)^H \mathbf{D} \exp(-\mathbf{W}z) = \mathbf{R}^{-1/2}(\mathbf{R}^{1/2}\mathbf{T}^H) e^{\Lambda z}(\mathbf{TR}^{1/2})\mathbf{R}^{1/2}\mathbf{D}\mathbf{R}^{1/2}(\mathbf{R}^{1/2}\mathbf{T}^H)e^{-\Lambda z}(\mathbf{TR}^{1/2})\mathbf{R}^{-1/2}$$

$$= \mathbf{T}^H e^{\Lambda z}\mathbf{TET}^H e^{-\Lambda z}\mathbf{T},$$

where $\mathbf{E}$ is a diagonal matrix with positive, real entries, even if one or more of the diagonal entries of $\mathbf{R}$ is negative (They cannot be zero by hypothesis.). It follows that $\mathbf{S} = \mathbf{TET}^H$ is symmetric, positive definite, and that $\mathbf{P} = e^{\Lambda z}\mathbf{S}e^{-\Lambda z}$, as a similarity transform of $\mathbf{S}$, has positive, real eigenvalues, $\lambda_1, \lambda_2, \ldots, \lambda_N$. In addition, the column vectors, $v_1, v_2, \ldots, v_N$, of $e^{\Lambda z}\mathbf{T}$ are eigenvectors of $\mathbf{P}$ that span $C^N$. We note the orthogonality of these vectors, since $(e^{\Lambda z}\mathbf{T})(e^{\Lambda z}\mathbf{T})^H = e^{2\Lambda z}$. Using the fact that the $v$'s span $C^N$, we find constant $\alpha_1, \alpha_2, \ldots, \alpha_N$ such that

$$\mathbf{T}y = \sum_i \alpha_i v_i,$$

The theorem follows since

$$\begin{aligned}
y^H \exp(\mathbf{W}z)^H \mathbf{D} \exp(-\mathbf{W}z)\, y &= \left(\sum_i \alpha_i v_i\right)^H \mathbf{P} \left(\sum_i \alpha_i v_i\right) \\
&= \left(\sum_i \alpha_i^* v_i^H\right)\left(\sum_i \alpha_i \lambda_i v_i\right) \\
&= \sum_i |\alpha_i|^2 \lambda_i \exp(2\lambda_i z) \\
&> 0.
\end{aligned}$$

■

# FIGURE CAPTIONS

*Figure 1*: Conceptual framework for pragmatic information. Details are explained in the text.

*Figure 2*: Time course of relative frequencies of various "replicators" evolving as described in the text, Equation (7), using the parameter matrix given in the text.

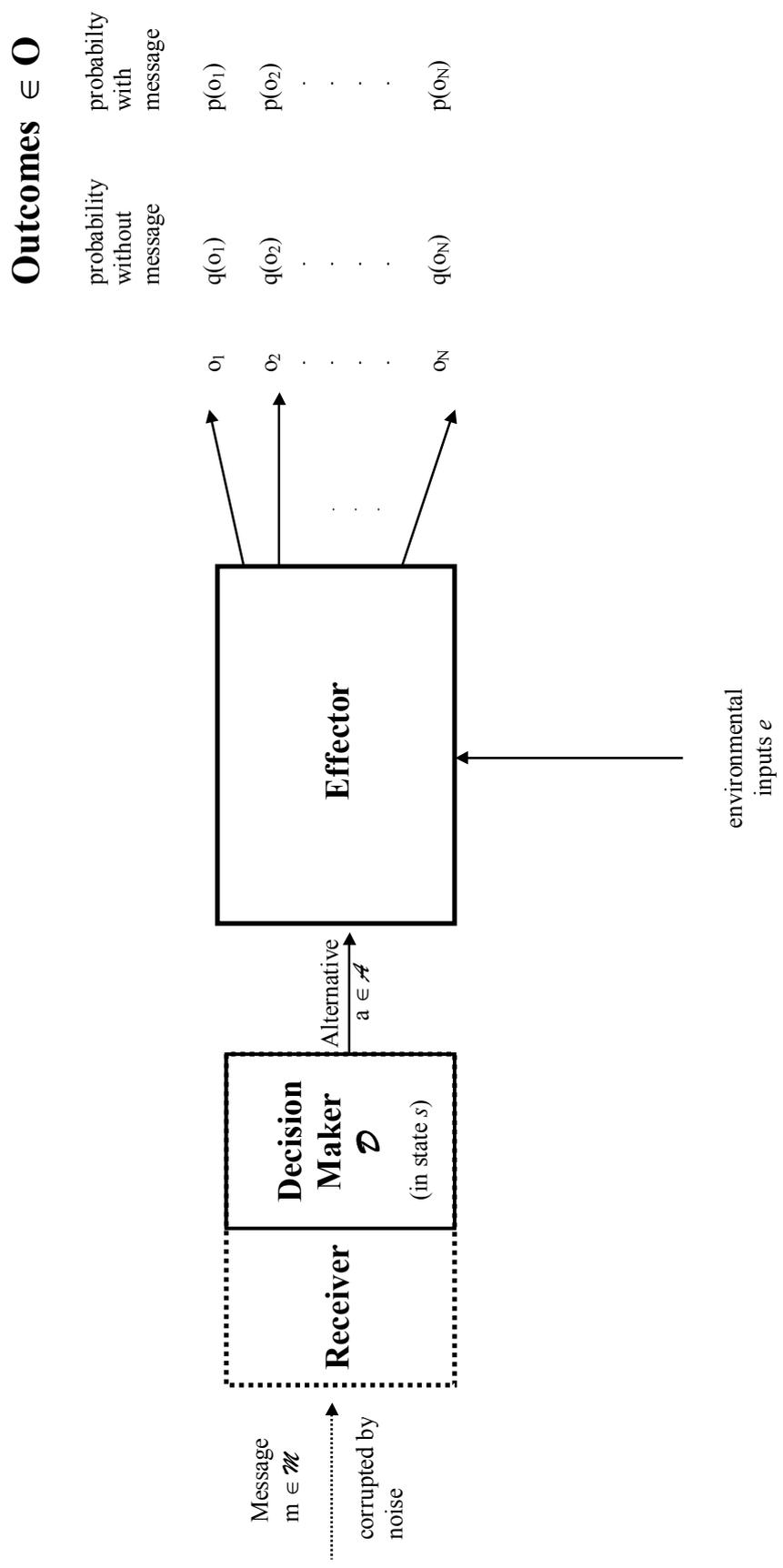

Figure 1

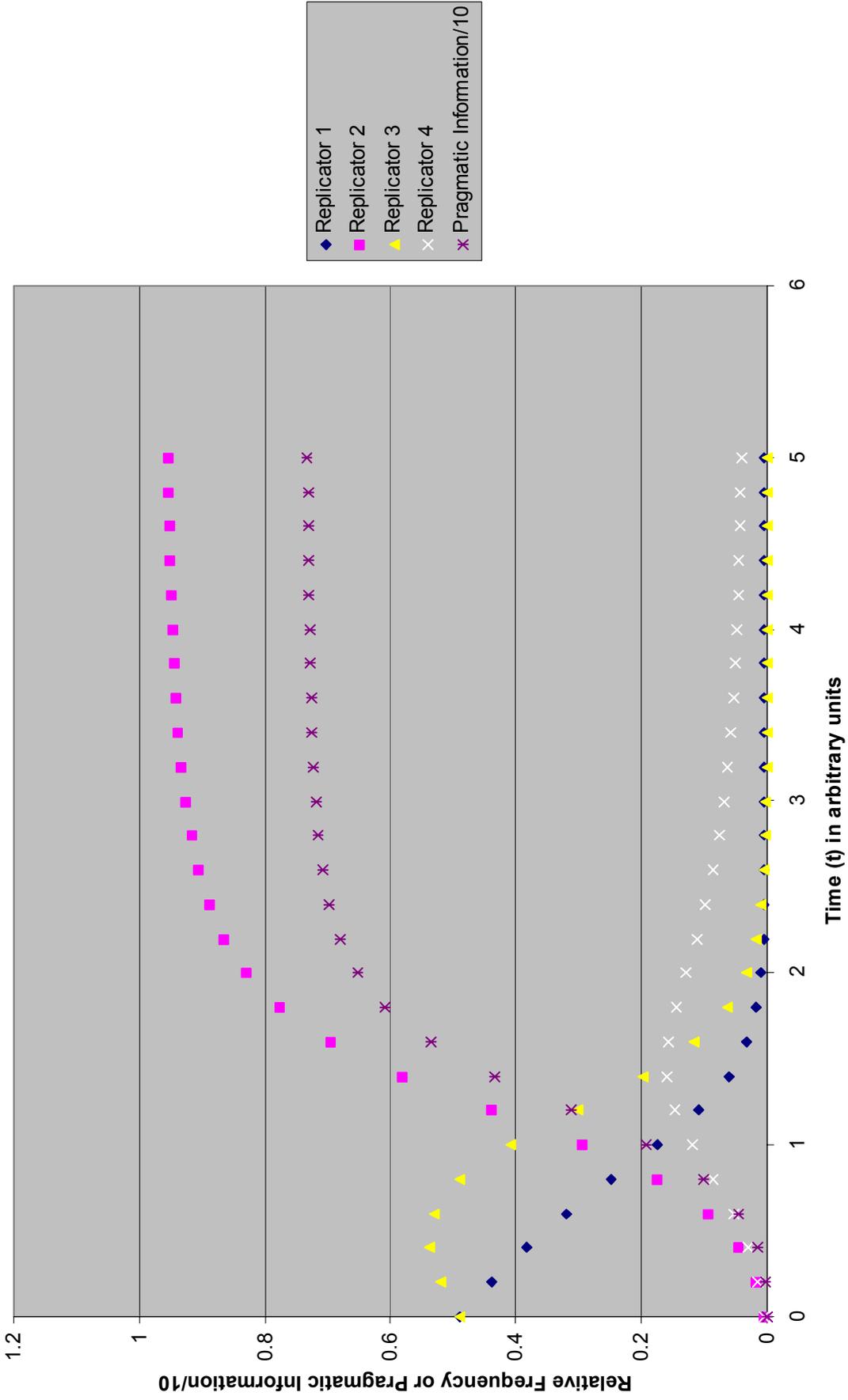

Figure 2